\documentclass[aps,prc,twocolumn,showpacs]{revtex4}
\usepackage{ulem}
\usepackage{color}
\usepackage{graphicx}
\usepackage{epsfig}
\usepackage{bm}
\def\nn {\nonumber}

\newcommand{\be}{\begin{equation}}
\newcommand{\ee}{\end{equation}}
\newcommand{\bea}{\begin{eqnarray}}
\newcommand{\eea}{\end{eqnarray}}

\newcommand{\om}{\omega}

\newcommand{\ov}{\overline}

\newcommand{\ds}{\partial \!\!\! /}

\newcommand{\bk}{\bm k}

\newcommand{\bq}{\bm q}
\newcommand{\bl}{\bm l}
\newcommand{\bu}{\bm u}

\newcommand{\del}{\partial}

\begin{document}

\title{Shear and Bulk Viscosities of Quark Matter from Quark-Meson Fluctuations 
in the Nambu--Jona-Lasinio model}
\author{Sabyasachi Ghosh$^1$,  Thiago C. Peixoto$^{1}$, Victor Roy$^2$, 
Fernando E. Serna$^{1}$, and Gast\~ao Krein$^{1}$}
\affiliation{$^{1}$Instituto de F\'{\i}sica Te\'orica, Universidade Estadual Paulista, 
Rua Dr. Bento Teobaldo Ferraz, 271 - Bloco II, 01140-070 S\~ao Paulo, SP, Brazil}
\affiliation{$^2$Institut f\"ur Theoretische Physik, Johann Wolfgang Goethe-Universit\"at, 
Max-von-Laue-Str. 1, D-60438 Frankfurt am Main, Germany}

\begin{abstract}
We have calculated the temperature dependence of shear $\eta$ and bulk $\zeta$ viscosities 
of quark matter due to quark-meson fluctuations. The quark thermal width originating from 
quantum fluctuations of quark-$\pi$ and quark-$\sigma$ loops at finite temperature is calculated 
with the formalism of real-time thermal field theory. Temperature-dependent 
constituent-quark and meson masses, and quark-meson couplings are obtained in 
the Nambu--Jona-Lasinio model. We found a non-trivial 
influence of the temperature-dependent masses and couplings on the Landau-cut structure of 
the quark self-energy. Our results for the ratios $\eta/s$ and $\zeta/s$, where $s$ is 
the entropy density (also determined in the Nambu--Jona-Lasinio model in the quasi-particle
approximation), are in fair agreement with results of the literature obtained from different 
models and techniques. In particular, our result for $\eta/s$ has a minimum very close to 
the conjectured AdS/CFT lower bound, $\eta/s = 1/4\pi$.  
\end{abstract}

\maketitle

% % % % % % % % % % % % % % % % % % % % % % % % % % % % % % % % % % % % % % % % % % % 
\section{Introduction}
\label{sec:intro}

A high-temperature, weakly interacting medium is naturally expected to be produced 
in heavy ion collision (HIC) experiments at high energies because the constituents 
of the medium~(quarks and gluons) should become almost free according to the asymptotic 
freedom property of quantum chromodynamics (QCD). However, the experimental finding of 
elliptic flow in HICs at the Relativistic Heavy Ion Collider (RHIC) leads to the 
interpretation that the medium produced in the collisions is actually a strongly 
interacting liquid, instead of a weakly interacting gas. This interpretation comes, 
in first place, from the ability of hydrodynamics to describe the RHIC data with a 
small value of the shear viscosity over entropy density ratio, 
$\eta/s$~\cite{Romatschke1,Romatschke2,Heinz1,Heinz2,Roy:2012jb,Heinz:2011kt,
Niemi:2012ry,Schenke:2011bn}. The shear viscosity coefficient~$\eta$ of a medium 
represents the ability of its constituents to transfer momentum over a distance 
comparable to their mean free path. When the inter-particle coupling is strong, 
momentum transfer takes place easily and, therefore, the shear viscosity of the 
matter is small. Several theoretical studies have estimated the ratio $\eta/s$, 
at weak and strong couplings and for different temperature regimes {\textemdash} a set 
of such studies is represented by Refs.~\cite{{AMY},{Arnold:2006fz},{Dobado1},{Dobado2},
{Muronga},{Nakamura},{Csernai},{Gyulassy},{Meyer_eta},{Nakano},
{Nicola},{Xu1},{Kapusta:2008vb},{Xu2},{Itakura},{Greco1},{Greco2},{Redlich_NPA},
{Chen2},{Purnendu},{Krein},{Weise1},{SSS},{Plumari},{Sarkar},{Cassing},{G_CAPSS},
{Weise2},{Ghosh_piN},{Sadooghi},{SGhosh},{HM},{Kinkar1}}. For earlier studies, 
see e.g.~Refs.~\cite{{Gavin},{Prakash},{Zhuang},{Hufner}}. An interesting feature 
of the results in Refs.~\cite{Csernai,Gyulassy,Kapusta:2008vb,Purnendu} is that 
$\eta/s$ reaches a minimum in the vicinity of a phase transition; the smallness 
of this minimum is significant in connection with a conjectured lower bound, 
$\eta/s = 1/4\pi$, known as the KSS bound, obtained in the context of AdS/CFT 
correspondence~\cite{KSS}.     

Similar to the shear viscosity, another transport coefficient is the bulk
viscosity, $\zeta$. It is defined as the proportionality constant between the 
non-zero trace of the viscous stress tensor to the divergence of the fluid velocity,
and usually it appears associated with processes accompanied by a change in fluid 
volume or density. The bulk viscosity has received much less attention 
than the shear viscosity in hydrodynamical simulations because its numerical value 
is assumed to be very small, as it is directly proportional to the trace of the 
energy-momentum tensor which generally vanishes for conformally symmetric matter. 
However, lattice QCD simulations~\cite{Lat1,Lat2} have shown that the trace of 
the energy momentum tensor of hot and dense QCD might be large near the QCD phase 
transition, which in turn indicates a non-zero, and possibly large
value of $\zeta$ as well as of $\zeta/s$ near the transition temperature. In the 
framework of pure-gauge lattice QCD, $\zeta/s$ near the transition temperature is 
estimated in Ref.~\cite{Meyer_zeta}. Analytical calculations employing different 
techniques and models have been used for estimating~$\zeta$ of strongly interacting 
matter, a list of references are~\cite{{Redlich_NPA},{Cassing},{Pratt},{Tuchin},
{Karsch},{Noronha},{Redlich_PRC},{Dobado_zeta1},{Vinod},{Dobado_zeta2},{Sarkar2},
{Defu},{Kinkar2}}; some of those~\cite{Pratt,Tuchin,Karsch} indicate a divergent 
behavior for $\zeta$ near the transition temperature.

In this work we employ the two-flavor Nambu--Jona-Lasinio (NJL) model~\cite{NJL} 
to evaluate the temperature dependence of the shear and bulk viscosities in the vicinity
of the crossover temperature. The NJL model is a very useful model for studying many 
aspects of the chiral structure of QCD in vacuum and at finite temperature and baryon 
density~\cite{NJL-reviews}. The model has also brought a great deal of insight into 
the problem of viscosities of strongly interacting matter~\cite{Redlich_NPA,Zhuang,Cassing,
G_CAPSS,Weise2,Kinkar1,Kinkar2}. A practical and transparent way to evaluate the contributions 
of quark-meson fluctuations to viscosities is to evaluate one-loop quark self-energy 
diagrams to obtain the quark relaxation time or, equivalently, the quark thermal 
width~\cite{G_CAPSS,Weise2}. Adopting this method, we have first analyzed the detailed 
Landau cut structures of the quark self-energy for quark-$\pi$ and quark-$\sigma$ loops,
where the temperature dependence of the quark and meson masses and their couplings
play an important role for its on-shell
contributions. The present work explicitly demonstrates the nontrivial contributions 
of the temperature dependence and momentum dependence of the quark thermal width 
to the viscosities, in the kinematic domains where the quark pole remains within the 
regions of the Landau cut. The new aspect explored in the present work relates to 
the contributions of quark-$\sigma$ and quark-$\pi$ fluctuations
%scalar~($\sigma$) and pseudoscalar~($\pi$) mesonic fluctuations 
to $\eta$ and $\zeta$, which become the main sources of dissipation beyond temperatures 
where pion decay into on-shell quark-antiquark pairs becomes possible. 
This temperature is commonly known as Mott temperature ($T_M$) above
which the threshold condition of pion dissociation i.e. $m_\pi(T) > 2 M_Q(T)$
is satisfied.
%Since the NJL model is not confining, this happens at the Mott 
%temperature $T_M$, which is defined by the condition $m_\pi(T_M) > 2 M_Q(T_M)$, 
%where $M_Q(T)$ is the temperature-dependent constituent quark mass. 
In a very recent work~\cite{LKW}, the NJL model was
used to investigate the role played by quark-meson fluctuations on the shear viscosity.
Our work is complementary to that one, as we use a different formalism to evaluate 
viscosities, consider also the bulk viscosity, use different quark-$\pi$ and quark-$\sigma$ 
couplings reflecting dynamical chiral symmetry breaking, and make a detailed 
study of the Landau cut structure of the quark self-energy. The paper is organized as 
follows: in the next section, we address the formalism of computing shear and bulk viscosities 
in terms of the quark thermal width, which is deducted from the quark self-energy diagram 
in the framework of real-time thermal field theory (RTF). Next, in results section, 
the analytic structure of the quark self-energy is rigorously discussed and its 
contribution to shear
and bulk viscosity coefficients of quark matter is also discussed. A summary
and conclusion is presented in the last section.

% % % % % % % % % % % % % % % % % % % % % % % % % % % % % % % % % % % % % % % % % % % 
\section{Formalism}
\label{sec:formal}

In the Kubo formalism~\cite{Zubarev,Kubo}, the shear and bulk viscosities are defined 
%in terms of the zero-frequency and zero-momentum limit of 
in the Lehmann spectral 
representation of the two-point correlation functions of operators involving the 
components of the energy-momentum tensor $T_{\mu\nu}$~\cite{Hosoya}:
\be
\left(
\begin{array}{c}
\eta \\[0.2true cm]
\zeta
\end{array}
\right) = \lim_{q_0 , |\bq| \rightarrow 0^+}  \; \frac{1}{q^0}
\left(
\begin{array}{c}
\frac{1}{20}A_\eta(q^0,\bq) \\[0.2true cm]
\frac{1}{2} A_\zeta(q^0,\bq)
\end{array}
\right)
\label{eta-zeta}
\ee
with the spectral functions $A_\eta(q)$ and $A_\zeta(q)$ given by
\bea
A_\eta(q) &=& \int d^4x \, e^{i q\cdot x} \,
\langle[\pi^{ij}(x),\pi^{ij}(0)]\rangle_\beta ,
\label{A_eta} \\
A_\zeta(q) &=& \int d^4x \, e^{i q \cdot x} \,
\langle[{\cal P}(x),{\cal P}(0)]\rangle_\beta ,
\label{A_zeta}
\eea
where
\bea
\pi^{ij}(x) &=& T^{ij}(x) - \frac{1}{3} \delta^{ij} T^{k}_{\,k}, 
\\[0.3true cm]
{\cal P}(x) &=& -\frac{1}{3} T^{i}_{\,i}(x) - c^2_s \, T^{00}(x) ~.
\eea
In the above equations, $c^2_s$ is the sound velocity 
and $\langle (..)\rangle_\beta$ denotes an appropriate thermal average.
%{\textemdash} in the present paper we use the Real-time Thermal Field-theory 
%(RTF) formalism~\cite{LeBellac}.  
We note here that the second term in the expression for 
${\cal P}(x)$ is necessary to account for energy conservation in a quasi-particle 
description of the medium, an approximation we use in the present paper {\textemdash}
a clear discussion on this issue can be found in Ref.~\cite{Arnold:2006fz}.

We use the NJL model to obtain the above correlation functions. The Lagrangian 
density of the model for $u$ and $d$ flavors is of the form
\be
{\cal L} = {\ov \psi}(i\ds-m_Q)\psi
+G \left[({\ov \psi}\psi)^2 + ({\ov \psi}i\gamma^5{\bm\tau}\psi)^2\right],
\label{LNJL}
\ee
where $m_Q = (m_u,m_d)$ is the current quark mass matrix and 
${\bm \tau} = (\tau^1,\tau^2,\tau^3)$ are the flavor Pauli matrices. The energy-momentum
tensor is given in terms of the Lagrangian density by
\be
T^{\mu\nu} = - g^{\mu\nu} {\cal L}
+ \frac{\del {\cal L}}{\del(\del_\mu\psi)}\del^\nu \psi
= g^{\mu\nu} {\cal L} + i{\ov \psi}\gamma^\mu\del^\nu\psi.
\ee
The model will be treated in the quasi-particle approximation or, equivalently, in 
the leading-order approximation in the $1/N_c$ expansion, where $N_c = 3$ is the 
number of colors, which is also equivalent to the traditional Hartree 
approximation of many-body theory. In this approximation, the thermal spectral 
functions can be expressed in terms of the quark propagator. 

\begin{figure}[h]
\includegraphics[scale=0.5]{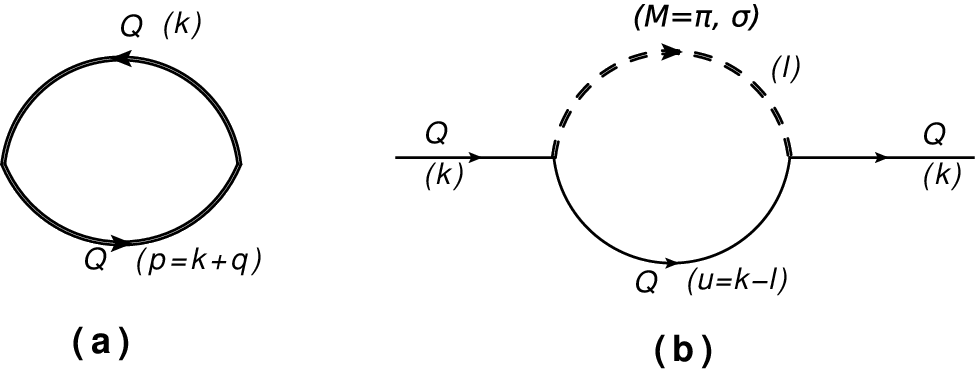}
\caption{The diagram {\bf (a)} represents a schematic one-loop diagram of the
quark correlator, which can be obtained from the two-point function of the viscous-stress
tensor for the quark constituents. The double-dashed lines
for the quark propagators indicate that they have some finite thermal
width which can be derived from the quark self-energy diagrams {\bf (b)}
for quark-meson loops (for $\pi, \sigma$ meson).}
\label{Shear_Q}
\end{figure}

We use the formalism of  real-time thermal field theory (RTF) to obtain the correlation 
functions. In RFT, the two point function of any field-theoretic operator has a 
$2 \times 2$ matrix structure reflecting the time ordering with respect to a
contour in the complex plane~\cite{LeBellac}.
The relevant matrix can be diagonalized 
in terms of a single analytic function, which determines completely the dynamics of 
the corresponding two-point function. In particular, the retarded correlation functions
needed for the evaluation of $A_\eta(q)$ and $A_\zeta(q)$ can be written in terms of
of the $11$ component of the corresponding two-point functions {\textemdash} see 
Ref.~\cite{G_Kubo} for details. For example, ignoring for the moment quark-meson fluctuations, 
$A_\eta(q)$ can be written as 
\be
A_\eta(q) = 2\, \tanh \left(\frac{\beta q_0}{2}\right){\rm Im} \, \Pi_{11}(q),
\label{R_bar_11}
\ee
with 
\be
\Pi_{11}(q) = i N_c N_f \int \frac{d^4k}{(2\pi)^4} \, N(q,k) 
D_{11}(k) D_{11}(q+k),
\label{P11}
\ee
where $D_{11}(q)$ is the scalar part of the $11$ component quark-propagator
matrix (in the zero width case):
\bea
D_{11}(k) &=&  \frac{-1}{k^2_0 - (\omega^k_Q)^2 + i\epsilon} \nn \\
&& \, -  2\pi i \omega^k_Q \, n_Q(\bk) \delta(k^2_0 - (\omega^k_Q)^2)~.
\label{D11}
\eea
The $n_Q(\bk)$ in Eq.~(\ref{D11}) is the Fermi-Dirac distribution 
\be
n_Q(\bk) = \frac{1}{1 + e^{\beta \omega^k_Q}},
\ee
with $\omega^k_Q = (\bk^2 + M^2_Q)^{1/2}$, and
\bea
N(q,k) &=& \frac{32}{3} k_0(k_0+q_0) \bk\cdot(\bk+\bq)  \nn \\
&& \, - 4 \left[\bk\cdot(\bk + \bq) + \frac{1}{3} \bk^2 (\bk+\bq)^2\right].
\label{N_qk}
\eea
%
%\be
%N(q,k)=-I_Qt^{\rho\sigma}_{\mn}{\rm Tr}[\gamma^\mu(q+k)^\nu(\qs+\ks+M_Q)
%\gamma_\rho k_\sigma(\ks +M_Q)]~.
%\label{N_qk}
%\ee
%
Fig.~\ref{Shear_Q}(a) shows this quark-quark loop diagram,
which can be considered as a schematic representation of 
shear viscosity coefficient at the zero
frequency and momentum limit.
The quark mass $M_Q$ is the solution of the gap equation  
\be
M_Q = m_Q + 4N_cN_fGM_Q\int\frac{d^3\bk}{(2\pi)^3}\frac{1 - 2 n_Q(\bk)}{\om^k_Q}.
\label{gap}
\ee
The temperature-independent part of the integral above is ultraviolet
divergent, while the temperature-dependent part, which contains the Fermi-Dirac
distribution function $n_Q(\bk)$, is finite. We use a ultraviolet cutoff $\Lambda$ to 
regularize the divergent integral, with $\Lambda$ fitted (together with the other parameters 
of the model, $m_Q$ and $G$) to obtain reasonable values for the quark condensate, 
pion leptonic decay constant and the pion mass in vacuum. The finite integral is integrated 
without a cutoff.

To proceed with the evaluation of the viscosities, we include the dissipative processes 
due to quark-meson fluctuations. The fluctuations introduce an imaginary part in the quark
self energy giving a thermal width $\Gamma_Q$ to the quarks; as it stands, with no finite 
imaginary part in the quark propagator, Eq.~(\ref{P11}) leads to divergent viscosities. 
Adding an on-shell, momentum-dependent thermal width $\Gamma_Q(k)$ in the quark propagator, 
one obtains~\cite{G_Kubo} for the shear viscosity the expression
\be
\eta = \frac{4N_cN_f}{15T} \int \frac{d^3\bk}{(2\pi)^3} \,  
\frac{\bk^4 \, n_Q(\bk) \left[1-n_Q(\bk)\right]}{(\om^k_Q)^2 \, \Gamma_Q(\bk)} .
\label{eta_final}
\ee
This expression for $\eta$ is identical to the one obtained from the standard 
relaxation time approximation~\cite{Gavin,Purnendu}. For bulk viscosity, from
Eq.~(\ref{A_zeta}) we obtain:
\bea
\zeta &=& \frac{4 N_c N_f}{T} \int \frac{d^3\bk}{(2\pi)^3} 
\, \frac{ n_Q(\bk) \left[1 - n_Q(\bk)\right]}{(\om^k_Q) ^2 \, \Gamma_Q(\bk)} 
\nn \\ 
&& \times \, \left[ \left(\frac{1}{3} - c_s^2\right) \bk^2 
- c_s^2 \frac{d}{d\beta^2} \left( \beta^2M^2_Q \right) \right]^2 ~,
\label{zeta_final}
\eea
which is also identical to the one obtained in Ref.~\cite{Purnendu} which uses
the relaxation time approximation. 

We evaluate the quark thermal width $\Gamma_Q(\bk)$ from the 
quark-meson ($QM$) loop contributions 
to the quark self-energy, pictorially represented in Fig~\ref{Shear_Q}(b). We need the
meson masses and quark-meson couplings. We assume the couplings of the mesons to the 
quarks are of Yukawa type~\cite{Quack_Klevansky}:
\be
{\cal L}_{QQ\pi} = i g_{QQ\pi} \overline{\psi}\gamma^5{\bm \tau \cdot \pi} \psi,
\label{LQQpi}
\ee
for the $\pi$ coupling, and 
\be
{\cal L}_{QQ\sigma} = g_{QQ\sigma} \overline{\psi}\sigma\psi,
\label{LQQsig}
\ee
for the $\sigma$ coupling. Note that the couplings $g_{QQ\sigma}$ and $g_{QQ\pi}$ are not
bare couplings; they incorporate the effects of dynamical chiral symmetry breaking and as
such are different from each other. At high temperatures, when chiral symmetry is restored,
they become equal to each other, as we discuss in the next section. The meson masses $m_M$ and 
quark-meson couplings $g_{QQM}$ are calculated from well-known expressions of the NJL 
model~\cite{NJL-reviews}:
\bea
&& \hspace{-0.5cm} 1 - 2 G \Pi_M(\omega^2=m^2_M) = 0, \label{m_M} \\[0.2true cm]
&& \hspace{-0.5cm} g^2_{QQM} = \left[\frac{\partial \Pi_M(\omega^2)}{\partial \omega^2} 
\right]^{-1}_{\omega^2 = m^2_M}  
\label{gQQM}, \\[0.2true cm]
&& \hspace{-0.5cm} \Pi_M(\omega^2) = 2N_C N_f \int \frac{d^3\bk}{(2\pi)^3} 
\frac{1-2 n_Q(\bk)}{\omega^k_Q} F_M(\omega^2), 
\label{PiM}
\eea
with
\bea
F_\pi(\omega^2) &=& \frac{(\omega^k_Q)^2}{(\omega^k_Q)^2 - \omega^2/4}, 
\label{Fpi} \\[0.2true cm]
F_\sigma(\omega^2) &=& \frac{(\omega^k_Q)^2 - M^2_Q}{(\omega^k_Q)^2 - \omega^2/4},
\label{Fsigma} 
\eea
where $M_Q$ is the solution of the gap equation in Eq.~(\ref{gap}). The integrals in
Eq.~(\ref{PiM}) are to be understood as principal-value integrals when $\omega^2 > 4 M^2_Q$. 
Like with the gap equation, the temperature independent part of the integral for
$\Pi_M(\omega^2)$ is divergent, and we use the same cutoff $\Lambda$ as in the gap 
equation.

Given the meson masses and quark-meson couplings, the quark thermal width is obtained 
from the imaginary part of the loop diagrams shown in Fig.~\ref{Shear_Q} (b). The off-shell
quark self-energy contains four branch cuts on the energy axis, but on-shell only the
Landau cut contributes~\cite{G_Kubo} and the result can be written as
\bea
\Gamma_Q &=& \sum_{M=\pi,\sigma} \Gamma_{Q(QM)} \nn \\
&=&  \biggl[\int\frac{d^3\bl}{(2\pi)^3}\, \delta(k^0 +\om^l_Q  - \om^u_M)\,
\frac{n_Q(\bl) + n_M(\bu)}{4 \om^l_Q \om^u_M} \nn \\[0.2true cm]
&& \times \, L_{Q(QM)}(k^0,\bk;l^0=-\om^l_Q,\bl) \biggr]_{k^0 = \omega^k_Q},
\label{gm_Q}
\eea
where $\bu = \bk - \bl$, $n_M(\bk)$ is the Bose-Einstein distribution  
\be
n_M(\bk) = \frac{1}{e^{\beta\om^k_M} - 1},
\ee
with $\om^k_M = (\bk^2 + m^2_M)^{1/2}$, and
\bea
L_{Q(Q\pi)}(k,l) &=& 3 \, \frac{4g_{QQ\pi}^2}{M_Q} \, \left( M_Q^2 - k \cdot l \right),
\\[0.2true cm]
L_{Q(Q\sigma)}(k,l) &=& \frac{4g_{QQ\sigma}^2}{M_Q} \, \left(M_Q^2 + k\cdot l\right).
\eea
Note that use of an on-shell quark thermal width is consistent with the 
quasi-particle approximation we are using to describe the system.

The last input needed is the sound velocity $c_s$, which is required for the evaluation of the 
bulk viscosity in Eq.~(\ref{zeta_final}). It can be obtained from the pressure $P$ and energy 
density $\epsilon$ as:
\be
c_s^2=\left(\frac{\del P}{\del \epsilon}\right)_{s}=
\frac{s}{c_V},
\label{cs2}
\ee
where $s$ is the entropy density and $c_V$ the specific heat at constant volume:
\be
s   = \frac{\epsilon + P}{T}, \hspace{0.75cm}
c_V = T\left(\frac{\del s}{\del T}\right)_{V}.
\label{s_T}
\ee
The pressure $P$ and energy density $\epsilon$ are given in the quasi-particle approximation
to the NJL model as
\be
\left(
\begin{array}{c}
P \\
\epsilon 
\end{array}
\right)
= 4 N_cN_f 
\int \frac{d^3\bk}{(2\pi)^3} \, n_Q(\bk)
\left(
\begin{array}{c}
\bk^2/3\om^k_Q \\
\om^k_Q
\end{array}
\right).
\label{P_e}
\ee
We note that the use of the quasi-particle approximation for the thermodynamic quantities
is consistent with the large $N_c$ counting~\cite{LKW}. Inclusion of meson loops 
contributions requires care with respect to thermodynamic consistency. 

% % % % % % % % % % % % % % % % % % % % % % % % % % % % % % % % % % % % % % % % % % % 
\section{Numerical results and discussion}
\label{sec:num}

The free parameters in the NJL model are the current quark mass $m_Q$, the coupling $G$ 
and the cutoff mass $\Lambda$ that is required to regularize vacuum loop diagrams. They
are fixed to obtain reasonable vacuum values for the quark condensate $\langle \bar uu\rangle$, 
pion leptonic decay constant $f_\pi$ and and the pion mass $m_\pi$. With $m_Q = m_u = m_d = 5$~MeV, 
$G\Lambda^2 = 2.14$ and $\Lambda = 653$~MeV, one obtains: $\langle \bar u u\rangle = 
(-252~\rm{MeV})^3$, $f_\pi = 94$~MeV and $m_\pi = 142$~MeV. The vacuum value of the 
constituent quark mass is $M_Q = M_u = M_d = 328$~MeV mass and of the $\sigma$ mass is
$m_\sigma = 663$~MeV.

\begin{figure}[t]
\includegraphics[scale=0.35]{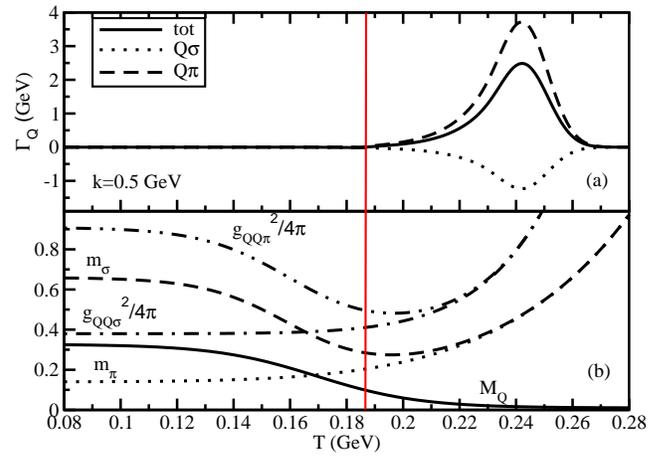}
\caption{(Color on-line) (a) Temperature dependence of $\Gamma_Q$ (solid line) 
for $k = 0.5$~GeV and the individual contribution from $\pi-$ (dashed line) and $\sigma-$quark 
loops (dotted line). (b) Temperature dependence of quark mass (solid line),
the pion (dotted line), and the sigma (dashed line) meson masses and
quark-meson couplings ($g^2_{QQ\pi}/4\pi$: dash-dotted line
and $g^2_{QQ\sigma}/4\pi$: dash-double-dotted line).
The straight vertical red line denotes the Mott temperature 
($T_M=0.187$ GeV).}
\label{g_t_T}
\end{figure}

We start our numerical discussion from on-shell quark 
thermal width $\Gamma_Q(\bk,T)$, which we present in the 
top panel of Fig.~(\ref{g_t_T}) for a specific value of the momentum, $\bk = 0.5$~GeV. Also shown 
are the individual contributions from the $\pi-$ and $\sigma-$quark loops. In the lower panel 
of the figure, we show the temperature dependences of the quark-meson couplings and quark, 
$\pi$ and $\sigma$ masses. The straight vertical red line in the figure denotes the Mott 
temperature $T_M$, at which $m_\pi(T_M)=2M_Q(T_M)$. For the parameter values used 
in the present paper,  $T_M=0.187$~GeV. Here one should notice that the numerical strengths 
of the pion and sigma modes are opposite in sign~\cite{Quack_Klevansky,Weise2}, but the 
magnitude of the pion mode is approximately three times larger than the magnitude of the 
sigma mode in the temperature range $T>T_M$. In the lower panel of the figure, one should 
notice that at sufficiently high temperatures, the $g_{QQ\pi} \simeq g_{QQ\sigma}$
as well as $m_\sigma \simeq m_\pi$, a feature 
due to the chiral symmetry restoration at high temperatures.

%%%%%%%% Figure 2

\begin{figure}[h]
\includegraphics[scale=0.35]{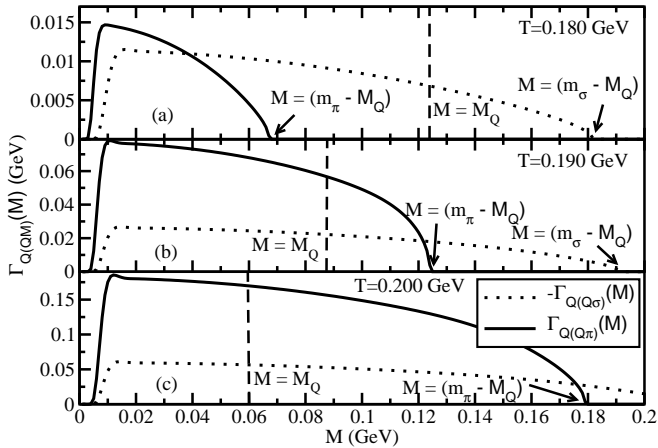}
\caption{Invariant mass distribution of the quark thermal width
from the $Q\pi$ (solid line) and $Q\sigma$ (dotted line) loops for 
$|\bk| = 500$~MeV for temperatures (a)~$T=0.180$~GeV, (b)~$0.190$~GeV, 
and (c)~$0.200$~GeV. The curve for $\Gamma_{Q(Q\sigma)}(M)$ as been multiplied 
by $-1$.}
\label{g_M_sigma}
\end{figure}

It is instructive to examine the invariant-mass $M$ dependence of the individual contributions 
$\Gamma_{Q(Q\pi)}$ and $\Gamma_{Q(Q\sigma)}$ to $\Gamma_Q$; $M$ is given by $M^2 = (k^0)^2 -\bk^2$. 
Results for $\bk = 0.5$~GeV and three different temperatures are shown in Fig.~\ref{g_M_sigma}.
The Landau-cut regions for quark-pion $(Q\pi)$ and quark-sigma $(Q\sigma)$ 
loops of the quark self-energy are respectively 
$0<M<|m_\pi -M_Q|$ and $0<M<|m_\sigma -M_Q|$; they are identified by the nonzero values of 
the corresponding $\Gamma_{Q{QM}}(M)$'s, as indicated in the figure. Clearly, the Landau-cut 
is very sensitive to the temperature as it is determined by the temperature-dependent 
quark mass $M_Q$ and meson masses $m_\pi$ and $m_\sigma$.  The strengths of the 
$\Gamma_{Q{QM}}$ are essentially controlled by the quark-meson coupling constants; 
please note the different scales in the vertical axes of the different panels. Also, while
the threshold condition ($m_\sigma-2M_Q\geq 0$) for the $Q\sigma$ loop is satisfied 
for all temperatures, the corresponding threshold condition for the $Q\pi$ loop 
($m_\pi-2M_Q\geq 0$) is only satisfied beyond the Mott temperature $T_M=0.187$~GeV. 
This last feature explains the fact that the quark-mass pole, indicated by the vertical dashed
lines in the figure, lies outside of the Landau cut for $Q\pi$ loop at $T=0.180$~GeV.

%%%%%%%% Figure 3

\begin{figure}[t]
\includegraphics[scale=0.35]{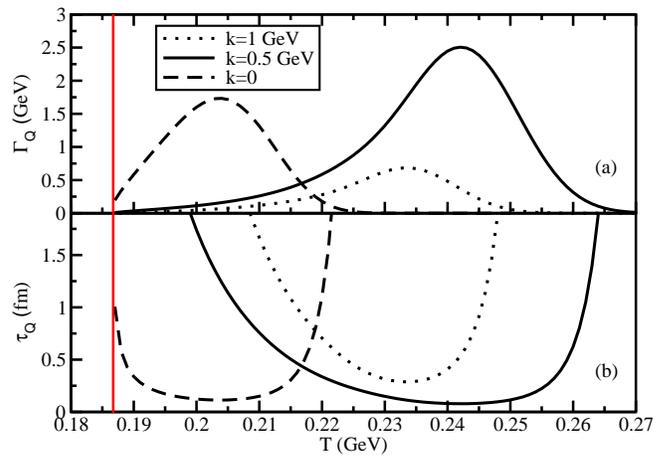}
\caption{(Color on-line) (a) Temperature dependence of the quark thermal width $\Gamma_Q$ 
and (b) collisional time $\tau_Q = 1/\Gamma_Q$ for three different values of momentum: 
$\bk=0$ (dashed line), $0.5$~GeV (solid line), and $1$~GeV (dotted line). The vertical 
red line indicates the Mott temperature. 
}
\label{g_T_k}
\end{figure}

Next we examine the temperature dependence of $\Gamma_Q$ and of it inverse, the collisional 
time $\tau_Q = 1/\Gamma_Q$, for three different values of momentum; the results are shown
in Fig.~\ref{g_T_k}. The peak position and strength of the temperature dependence of 
$\Gamma_Q$ is strongly momentum dependent, a feature that reflects the absorption and emission
processes of the quark interacting with mesonic modes; recall that $\Gamma_{QQM}$ physically 
means the on-shell probability of forward and inverse scattering between a quark $Q$ and the 
mesonic modes $M$~\cite{Weldon}. In forward scattering, $Q$ may be transformed to a thermalized 
$M$ by absorbing a thermalized anti-quark ${\bar Q}$, while in the inverse process, an 
off-equilibrated $Q$ may be regenerated via $M\rightarrow Q{\bar Q}$ dissociation. We should 
note as well that $\Gamma_Q$ decreases and ultimately goes to zero for large
temperatures; the temperature, for which it tends to be numerically zero, 
depends of course on $\bk$ and 
is determined by the temperature-dependent quark and meson masses and couplings. The implication
of $\Gamma_Q$ decreasing with $T$ for large $T$ has the implication that $\eta$ and $\zeta$
will increase with $T$, and again, the value of $T$ where it starts to increase and the rate 
of increase depend on the parameters of the model.  

When considering the collisional time, shown in the lower panel of Fig.~\ref{g_T_k}, one can 
compare time scales of dissipation with the typical evolution time $\Delta\tau$ of the matter 
produced in a HIC, which is $\Delta\tau \simeq 1-10$~fm. One can clearly identify well-defined 
temperature ranges for which $\tau_Q < \Delta\tau$, i.e. for which the propagating quark in the 
medium suffers sufficient dissipation by scattering with the thermalized constituents of the medium. 

%%%%%%%% Figure 4

%%%%%%%%%  Figure 5
\begin{figure}[h]
\includegraphics[scale=0.35]{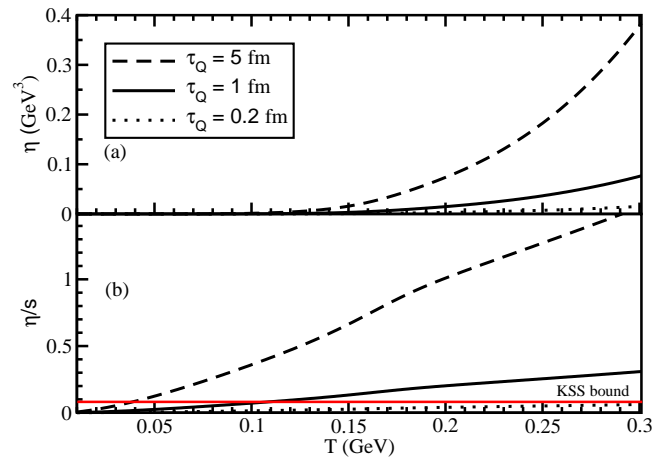}
\caption{(Color on-line) (a) Temperature dependence of $\eta$ and (b) $\eta/s$ for different
values of constant collisional time: $\tau_Q=5$~fm (dashed line), 
$1$~fm (solid line), and $0.2$~fm (dotted line). The straight horizontal red line indicates 
the KSS bound, $\eta/s = 1/4\pi$.}
\label{eta_T2}
\end{figure}

Before using the full temperature- and momentum-dependent $\Gamma_Q(k)$ in  
Eqs.~(\ref{eta_final}) and (\ref{zeta_final}), we have calculated the temperature dependence 
of $\eta$ and $\zeta$ for a constant value of $\tau_Q = 1/\Gamma_Q$. 
Although the temperature and momentum dependence of $\Gamma_Q$ will modify somewhat the 
results for the viscosities, the calculation with a constant $\Gamma_Q$ will bring insight 
regarding the integrands in~Eqs.~(\ref{eta_final}) and (\ref{zeta_final}) when $\Gamma_Q$ 
can be taken out of the integrals. In~Fig.~\ref{eta_T2} we show the corresponding results 
for $\eta$ and $\eta/s$, for different values of $\tau_Q$. While $\eta$ is a 
monotonically increasing function of $T$, the ratio $\eta/s$ exhibits two different 
rates of increase in two different temperature domains, which are associated with two 
different phases. 

%%%%%%%%  Figure 6
\begin{figure}[hb]
\includegraphics[scale=0.3]{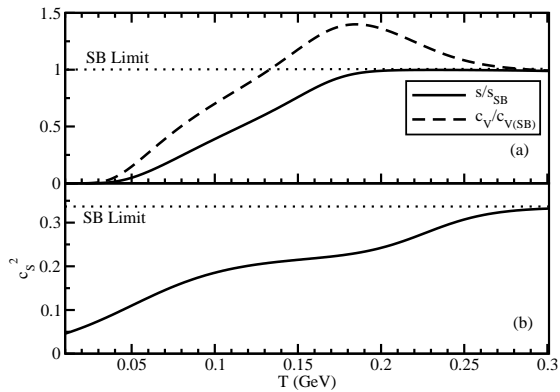}
\caption{(a) The temperature dependence of entropy density, normalized
by Stephan-Boltzmann (SB) limiting value $s_{SB}$, given in Eq.~(\ref{s_SB}). 
(b) The squared speed of sound $c_s^2$ as a function of the temperature. 
Straight horizontal dotted lines stand for the corresponding SB limits.}
\label{s_cs2_T}
\end{figure}

The observed change in the rate of increase of $\eta/s$ with the temperature can 
be understood from the temperature dependence of $s$, normalized by the 
Stephan-Boltzmann (SB) limiting value: 
\be
s_{SB}= 4 N_cN_f \left(\frac{7}{2}\right) \left(\frac{\pi^2}{90}\right)T^3.
\label{s_SB}
\ee
One observes in Fig.~\ref{s_cs2_T}(a) that the rate of increment in $s/s_{SB}$ mainly 
changes around $T\approx 0.175-185$ GeV; above this temperature, the entropy density of 
quark matter becomes almost identical with the SB limiting value, which is denoted by 
straight horizontal dotted line in the figure. It is important to note that we are taking 
into account the contributions of the quarks only in the expression for $s_{SB}$; including 
gluon degrees of freedom will increase $s_{SB}$ by a factor roughly equal to $1.75$ 
(for $N_f =2$). Being proportional to the slope of~$s$, the specific heat $c_V$ is 
magnifying the transition by exhibiting a smooth hump structure around 
$T\approx 0.175-185$ GeV as shown in Fig.~\ref{s_cs2_T}(a). Such a behavior of $c_V$ has 
been obtained also in other recent calculations employing the NJL model, as e.g. in 
Refs.~\cite{{Cassing},{Chatterjee:2011jd},{Bhattacharyya:2012rp}}. Note, however, that the 
maximum of $c_V$ is still well below the corresponding SB limit in QCD which, of course, 
includes gluon degrees of freedom and is thereby $1.75$ larger than the pure-quark SB limit. 
This is a welcome feature as lattice QCD simulations~\cite{Lat1} show no indication that the 
specific heat exceeds the QCD SB limit at any temperature. For calculating the bulk viscosity, 
we have obtained $c_s^2$ using Eq.~(\ref{cs2}); its temperature dependence is shown in 
Fig.~\ref{s_cs2_T}(b). Assuming the system is not very far from equilibrium, all 
thermodynamical quantities are obtained for non-interacting system of quark matter, 
although a finite (not zero) probability of quark-meson interaction has to be considered 
for getting a non-divergent values of transport coefficients.

\begin{figure}[h]
\includegraphics[scale=0.35]{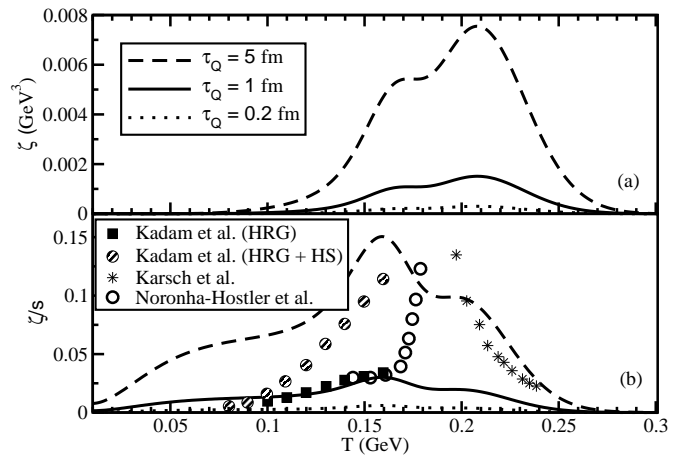}
\caption{(a) Temperature dependence of $\zeta$ and (b) $\zeta/s$ for different
values of constant $\tau_Q$. In the lower panel, also shown are results 
from the literature: Ref.~\cite{HM} (filled circles and squares), 
Ref.~\cite{Noronha} (open circles), Ref.~\cite{Karsch} (stars).}
\label{zeta_T2}
\end{figure}

\begin{figure}[htb]
\includegraphics[scale=0.3]{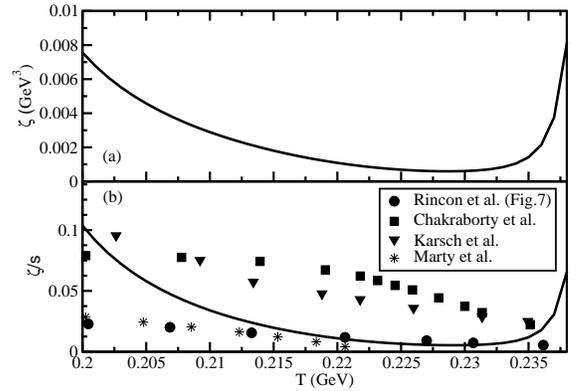}
\caption{(a) Temperature dependence of $\zeta$ and (b)~$\zeta/s$ using the full momentum 
dependence of the quark thermal width $\Gamma_Q(\bk)$. Comparison with results from the literature:  
Ref.~\cite{Dobado_zeta2} (circles), Ref.~\cite{Purnendu} (squares), 
Ref.~\cite{Karsch} (triangles), Ref.~\cite{Cassing} (stars).}
\label{zeta_T}
\end{figure}

Fig.~\ref{zeta_T2} shows results for the temperature dependence of $\zeta$ and $\zeta/s$,
for different constant values of $\tau_Q$. A~non-monotonic temperature dependence of 
$\zeta$ is obtained, with a distinctive two-peak structure at temperatures 
$T\approx 0.150-0.160$~GeV and $T\approx 0.210-0.220$~GeV. Similar double-peak structure for
$\zeta$ was obtained in Ref.~\cite{Dobado_zeta2}. Note that the first peak is located
at a temperature below the Mott temperature and will not be seen (Fig.~\ref{zeta_T}) 
when using the full temperature and momentum dependence of $\Gamma_Q(\bk,T)$. 

%%%%%%%%%  Figure 7

%%%    Figure 8
\begin{figure}[ht]
\includegraphics[scale=0.35]{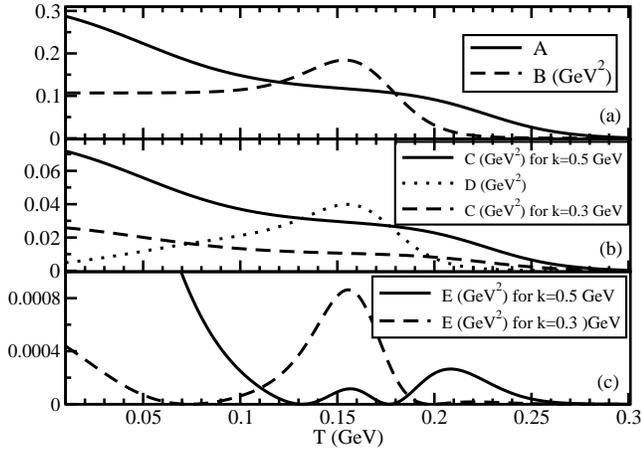}
\caption{Temperature dependence of the auxiliary quantities defined in the text. 
Panel~(a): $A=1/3-c_s^2$ (solid line) and $B = d(\beta^2M^2_Q)/d\beta^2$ (dashed line).
Panel~(b): $C=A \, \bk^2$, for $|\bk|=0.5$~GeV (solid line) and $|\bk|=0.3$~GeV (dashed line), 
$D=c_s^2B$ (dotted line). Panel~(c): $E=(C-D)^2$ for $|\bk|=0.5$~GeV (solid line)
and $|\bk|=0.3$~GeV (dashed line).}
\label{conformality}
\end{figure}

The double-peaked structure is clearly the result of a com\-pe\-ti\-tion be\-tween the two 
con\-for\-mal\-ity break\-ing terms in the integrand of Eq.~(\ref{zeta_final}), 
$1/3 - c_s^2$ and $d(\beta^2M^2_Q)/d\beta^2$: the first is associated with 
$c^2_s \neq 1/3$ and the second is associated with the bulk mass transport~\cite{Purnendu}. 
To get a graphical understanding of the competition, let us define auxiliary quantities 
containing these terms: $A = 1/3 - c_s^2$,  $B = d(\beta^2M^2_Q)/d\beta^2$, 
$C = A \, \bk^2$, $D = c_s^2 \, B$ and $E = (C-D)^2$. These quantities are plotted 
in the different panels in Fig.~\ref{conformality}.

The key feature here is that the $C$ (with any fixed momentum) and $D$ curves intersect each 
other at three different points; therefore, $E = (C-D)^2$ possesses three nodes (minima) 
and, of course, two maxima along the $T$-axis. These two maxima remain when $E = E(\bk)$ is 
integrated over $\bk$ in Eq.~(\ref{zeta_final}) and the two-peak structure is clearly understood
in these terms. Of course, if $M_Q$ is temperature independent, then one has only one peak, as
the lower peaks in Figs.~\ref{zeta_T2} and \ref{conformality}(c) will not appear. 
Likewise, as seen in the lower panel of Fig.~\ref{zeta_T2}, the ratio 
$\zeta/s$ also presents a non-monotonic behavior. For the sake of comparison, 
Fig.~\ref{zeta_T2}(b) also displays results from the literature.

\begin{figure}[htb]
\includegraphics[scale=0.35]{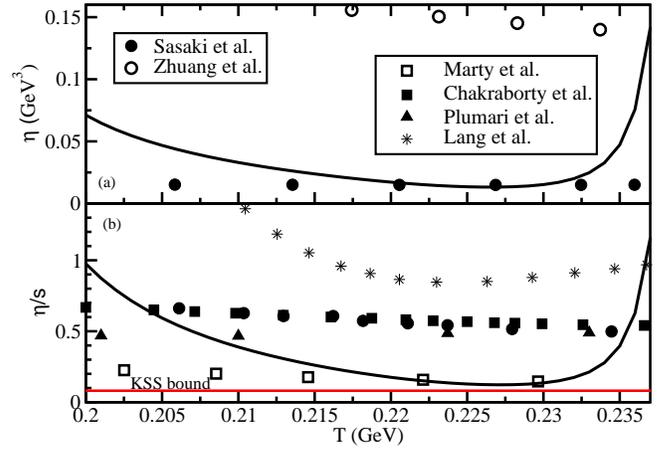}
\caption{(Color on-line) (a) Temperature dependence of $\eta$ and 
(b)~$\eta/s$ using the full momentum dependence of the quark thermal width $\Gamma_Q(\bk)$. 
Comparison with results from the literature: Ref.~\cite{Redlich_NPA} (filled circles),
Ref.~\cite{Zhuang} (open circles), Ref.~\cite{Cassing} (open squares), 
Ref.~\cite{Purnendu} (filled squares), Ref.~\cite{Plumari} (triangles),
Ref.~\cite{Weise2} (stars).}
\label{eta_T}
\end{figure}

Finally, we present results for the viscosities and viscosity to entropy density ratios 
with the full momentum dependence of the quark thermal width~$\Gamma_Q(\bk)$. It 
is important to reiterate that our results have physical significance only within 
a limited range of temperatures, approximately $T_M < T < 0.240$~GeV. The lower
limit is due to the fact that in the NJL model, when using the on-shell thermal 
width, finite viscosities are obtained for temperatures $T$ larger than the Mott 
temperature $T_M$ because $\Gamma_Q(k)$ is nonzero only for $T > T_M$ in this 
case. As discussed earlier, the off-shell quark self-energy contains four branch 
cuts on the energy axis, but on-shell only the Landau cut contributes. The upper 
limit is not a precise limit, but it is related to the lack of asymptotic freedom 
in the NJL model; it is known that the large temperature behavior of viscosities 
in QCD, particularly of bulk viscosity, is very different from the one derived 
from a non-gauge, contact-interaction model~\cite{Arnold:2006fz}. Given this 
proviso, we discuss next our results for the temperature dependence of $\eta$ 
and $\eta/s$, and $\zeta$ and $\zeta/s$; they are presented respectively 
in Figs.~\ref{eta_T} and ~\ref{zeta_T}. 

%%%% Figure 9

First of all, one notices in Figs.~\ref{eta_T} and ~\ref{zeta_T} that the use of the full 
temperature dependence of $\Gamma_Q(k)$ restricts the domain of temperatures where finite 
viscosities are obtained. Our numerical values of $\eta$ and $\eta/s$ within the 
temperature range $T_M < T < 0.240$~GeV, are quite close to some earlier estimates in
Refs.~\cite{{Redlich_NPA},{Zhuang},{Cassing},{Purnendu},{Plumari},{Weise2}}. 
Similarly, Fig.~(\ref{zeta_T}) shows fair agreement of our calculation of $\zeta/s$ with some earlier 
results of Refs.~\cite{{Dobado_zeta2},{Purnendu},{Karsch},{Cassing}}.

%%% Figure 10

The interesting two-peak structure in $\zeta$ shown in Fig.~(\ref{zeta_T2}) is not visible 
here because, as discussed above, the first peak is below the Mott temperature. In addition, 
the figures show that the viscosities have a rapid increase with temperature for $T \ge 0.235$~GeV. 
This is due to the fact that for high temperatures, 
the contributions from quark-meson fluctuations 
to $\Gamma_Q(k)$ decrease rapidly for large $T$, as shown in the top panel of Fig.~\ref{g_t_T}. 
At higher temperatures, quark-quark and quark-gluon scatterings will become more important 
than quark-meson scattering. To include such processes, one would probably need a model
that incorporates asymptotic freedom, as already mentioned in the previous paragraph. As 
our focus of interest in the present is on the contributions from quark-meson thermal 
fluctuations, we reserve for a future publication the investigation of the contributions
of quark-quark and quark-gluon processes to viscosities. 

%%% Figure 11
\begin{figure}[h]
\includegraphics[scale=0.3]{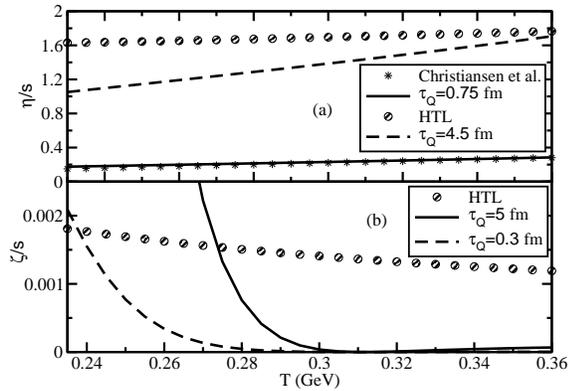}
\caption{$\eta/s$ (a) and $\zeta/s$ (b) at high temperature domain.
HTL (circles) and FRG (stars) estimations of $\eta/s$ by Arnold et al.~\cite{AMY} and
Christiansen et al.~\cite{Nills} are compared with $\eta/s$ of Eq.~(\ref{eta_final}) 
at $\tau_Q = 4.5$~fm (dashed line) and $\tau_Q = 0.75$~fm (solid line). 
While HTL (circles) estimation of $\eta/s$ by Arnold et al.~\cite{Arnold:2006fz} 
is compared with $\zeta/s$ of Eq.~(\ref{zeta_final}) at $\tau_Q = 0.3$~fm 
(dashed line) and $\tau_Q = 5$~fm (solid line).}
\label{eZta_T_ref}
\end{figure}

We conclude with a discussion of the high-temperature behavior of $\eta/s$ and $\zeta/s$ 
by contrasting results from the literature and what one would obtain from our 
Eqs.~(\ref{eta_final}) and (\ref{zeta_final}) when using finite values for the collisional 
time~$\tau_Q$. Fig.~(\ref{eZta_T_ref}) is showing results for these ratios for $T > 0.240$~GeV 
from hard thermal loop (HTL) calculations~\cite{AMY,Kapusta:2008vb,Csernai} and from a 
functional renormalization group (FRG) calculation for pure Yang-Mills theory~\cite{Nills}. 
For $\eta/s$ we note that the HTL and FRG results are markedly different. The FRG result is
almost identical to the one obtained from Eq.~(\ref{eta_final}) with $\tau_Q = 0.75$~fm are 
seem to match very well with our result shown in Fig.~(\ref{eta_T})(b) form temperatures around
$T = 0.23$~GeV. On the other hand, for $\zeta/s$ there are qualitative and quantitive difference
between the HTL results from those from Eq.~(\ref{zeta_final}) for the two values of
$\tau_Q$ used{\textemdash}there are no FRG results available for $\zeta/s$. The discrepancies 
are significant and clearly further studies are required.

% % % % % % % % % % % % % % % % % % % % % % % % % % % % % % % % % % % % % % % % % % %
\section{Summary and Perspectives} 
\label{sec:summ}

In summary, we have investigated the temperature dependence of the quark-matter shear 
$\eta$ and bulk $\zeta$ viscosities and their ratios to the entropy density $s$. The 
focus of interest was on the contributions from $Q\pi$ and $Q\sigma$ fluctuations 
to the on-shell quark thermal width $\Gamma_Q$, a crucial ingredient in the calculation 
of viscosities. We calculated $\Gamma_Q(\bk, T)$ from the imaginary part of the quark self-energy 
arising from $Q\pi$ and $Q\sigma$ loops using temperature-dependent quark and meson 
masses and quark-meson couplings. We have investigated in detail the Landau-cut structure of 
the quark self-energy using real-time thermal field-theory. The temperature dependence of 
masses and coupling constants were obtained in the NJL model in the quasi-particle approximation. 
The entropy density and the speed of sound, which is needed to calculate $\zeta$, were also 
obtained in the NJL model within the quasi-particle approximation. 

The temperature dependence of the masses and quark-meson couplings non-trivially influences
the Landau-cut structure of the quark self-energy. The quark thermal width, calculated from the
on-shell quark self-energy, becomes non-zero and non-negligible within a range of temperature 
and momentum where the quark pole is situated well inside the Landau-cut. When using
a temperature- and momentum-independent quark thermal width $\Gamma_Q$, a monotonically 
increasing shear viscosity as a function of temperature was obtained, whereas, the bulk 
viscosity exhibits a two-peak structure near the phase transition 
temperature, reflecting the competition between conformality breaking terms related 
with the speed of sound in the medium and the bulk mass transport. On the other hand, 
when the full temperature- and momentum-dependent thermal width is used, finite viscosities
are obtained only for temperatures larger than the Mott temperature~$T_M = 0.187$~GeV. 
In addition, for high temperatures, due to rapidly increasing mesonic masses and 
decreasing of the constituent quark mass, the probability for quark-meson fluctuations
become negligible and the viscosities increase considerable and other processes not
related to quark-meson processes take place. Between the two extremes, our results
for the $\eta/s$ and $\zeta/s$ are in fair agreement with results from the literature,
although marked differences can be observed. 

Future work includes the investigation of viscosities at high temperatures by the introduction
of quark-quark and quark-gluon processes. Of course, agreement with standard perturbative 
QCD results are expected when using interactions that incorporate the property of asymptotic
freedom of QCD. At low temperatures, when the medium is dominated by pions and low-mass 
resonances~\cite{Ghosh_piN}, the NJL model can be used to obtain pure meson-meson interactions, 
deduced e.g. via the bosonization technique of NJL model, and results are expected to be very 
close to the standard results for the viscosities of hadronic matter.

{\bf Acknowledgment:} 
Work partially financed by Fun\-da\-\c{c}\~ao de Amparo \`a Pesquisa do Estado de 
S\~ao Paulo - FAPESP, Grant Nos. 2009/50180-0 (G.K.), 2012/16766-0 (S.G.), and
2013/01907-0 (G.K.); Conselho Nacional de Desenvolvimento Cient\'{\i}fico e 
Tecnol\'ogico - CNPq, Grant No. 305894/2009-9 (G.K.). T.C.P and F.E.S. are supported 
by scholarships from Conselho Nacional de Desenvolvimento Cient\'{\i}fico e 
Tecnol\'ogico - CNPq. V.R. is supported by the Alexander von Humboldt Foundation. 
We thank  Dirk H. Rischke for a careful reading of an earlier version the manuscript 
and valuable comments.

\end{document}